\documentstyle[12pt]{article}
 
\newcommand \blackboardrrm{\mathchoice
{\rm I\kern-0.21 em{R}}{\rm I\kern-0.21 em{R}}
{\rm I\kern-0.19 em{R}}{\rm I\kern-0.19 em{R}}}

\newcommand \blackboardzrm{\mathchoice
{\rm Z\kern-0.32 em{Z}}{\rm Z\kern-0.32 em{Z}}
{\rm Z\kern-0.28 em{Z}}{\rm Z\kern-0.28 em{Z}}}

\newcommand \be  {\begin{equation}}
\newcommand \ee  {\end{equation}}

\renewcommand \d {{\mbox d}}

\begin{document}

\title{Logarithmic Corrections for Spin Glasses,  Percolation
and Lee-Yang Singularities in Six Dimensions}

\author{Juan J. Ruiz-Lorenzo\\[0.5em]
{\small  Departamento de F\'{\i}sica Te\'orica I}\\
{\small  Facultad de Ciencias F\'{\i}sicas}\\
{\small  Universidad Complutense de Madrid }\\
{\small   \ \  Ciudad Universitaria }\\[0.3em]
{\small   \ \  28040 Madrid (Spain) }\\[0.3em]
{\small   \tt ruiz@lattice.fis.ucm.es}\\[0.5em]
} 

\date{June 26, 1998}

\maketitle

\begin{abstract}

We study analytically the logarithmic corrections to the critical
exponents of the critical behavior of correlation length,
susceptibility and specific heat for the temperature and the
finite-size scaling behavior, for a generic $\phi^3$ theory at its
upper critical dimension (six).  We have also computed the leading
correction to scaling as a function of the lattice size. We
distinguish the obtained formulas to the following special cases:
percolation, Lee-Yang (LY) singularities and $m$-component spin glasses.
We have compared our results for the Ising spin glass case with
numerical simulations finding a very good agreement. Finally, and
using the results obtained for the Lee-Yang singularities in six
dimensions, we have computed the logarithmic corrections to the
singular part of the free energy for lattice animals in eight
dimensions.

\end{abstract}  

\thispagestyle{empty}
\newpage

\section{\protect\label{S_INT}Introduction}

One of the techniques commonly used in the study of statistical
systems is to perform numerical simulations focusing on finite size
effects.  The main tool of this approach is the knowledge of how some
observables diverge in the critical region as a function of the size
of the system instead of the usual formulas that express these
divergences as a function of the reduced temperature (or in reduced
probability in the case of percolation) or magnetic field. Moreover it is
possible to measure these finite size effects in experiments.

For statistical systems below their upper critical dimensions there
is an extensive literature on this subject~\cite{CARDY}.

The main goal of this work is to obtain the functional form of the
divergences, as functions of the reduced temperature as well as the
lattice size. We focus on observables commonly measured in numerical
simulations at the upper critical dimension, for a wide class of
systems like the vector spin glasses~\cite{BOOK},
percolation~\cite{STAUFFER} and Lee-Yang~\cite{YL,FISHERA} (LY)
singularities. A classical feature of the upper critical dimension is
that the critical behavior (which is described by the critical
exponents) is modified by logarithms. The logarithmic corrections to
the critical behavior of the susceptibility (spin glasses (in reduced
temperature) and percolation (in probability)) and correlation length
(for percolation, in probability) were computed in references
~\cite{PERC_FT} (percolation) and ~\cite{FISCH} (spin glasses). We
will use these previous results as check of our calculation.

Moreover, and using the mapping proposed by Parisi and Sourlas
\cite{PASOUR} between the actions which describe the Lee-Yang
singularities for the Ising model in $d$ dimensions and the lattice
animals in $d+2$ dimensions, we have been able to compute the
logarithmic correction to the singular part of the free energy for
lattice animals \cite{STAUFFER,HALU} at its upper critical dimension
(eight). To check the mapping (a further check) we have compared our
result for the LY singularities with that of reference \cite{LUBENSKY}
where originally was computed the logarithmic correction to the free
energy directly for lattice animals and with reference
\cite{L_ANIMALS} where it was checked using series expansions.

The understanding of these logarithms has  very important physical
applications. For instance, the $\phi^4$ theory in four dimensions
(that we denote as $\phi^4_4$) is trivial (the theory that we obtain
when the cut-off is sent to infinity is a free, non-interacting,
theory) because the logarithmic corrections produce a vanishing
renormalized coupling constant~\footnote{ 
        At the critical point, the
        renormalized constant, $g_R=(L/\xi)^d B$ 
        (where $\xi$ is the correlation length, 
        $L$ is the lattice size, $d$ is the dimension and 
        $B$ is the Binder cumulant), 
        of the four 
        dimensional Ising model drops 
        following a law: $g_R \propto 1/\log L$ \cite{4DDIS}.}. 

Another example where the knowledge of the logarithmic factors is very
important is the study of the uniaxial system with strong dipolar
forces. In this system the upper critical dimension is just three and
thereby, the theoretical predictions for the logarithmic corrections
have been checked experimentally, the agreement being very good
\cite{ZINN}.

In this note we mainly focus on the study of the logarithms in the field
theory description of spin glasses, a $\phi^3$ theory~\footnote{
        This theoretical description only holds in the paramagnetic
        phase. To study the spin glass phase we need to consider a $\phi^4$
        term that induces the breaking of the replica symmetry 
        in infinite dimensions \cite{PARISI}.}. 
The generic $\phi^3$ theory (i.e. the coupling is
a generic tensor $\lambda_{ijk}$) also describes a large set of
interesting statistical systems. We can cite, for instance, 
the $q$-states Potts model, percolation and  Lee-Yang singularities 
(described by one-component  $\phi^3$ theory with a purely 
imaginary coupling \cite{FISHERA}). 

The study of spin glasses in finite dimensions is another interesting
current research issue~\cite{PARISI}. It is very important to
understand if the strange and interesting properties of the Parisi
solution (which is believed to be exact in infinite dimensions) 
apply in finite
dimensions~\cite{PARISI}. In particular, the existence of a large
number (infinite) of pure states which organize in an ultrametric
fashion is an open problem in the current spin glass research.

There exist two analytical approaches that try to answer these
questions: The first one is the droplet model~\cite{DROPLET,NS} that
predicts that the spin glass phase is composed by one pure state (and
its inverse by flipping all spins). The underlying approximation is
the Migdal-Kadanoff one that is an approximate real space
renormalization group. The Migdal-Kadanoff technique is known to give
exact answers in one dimension and also that it lacks of predictive
power when the dimensionality grows. For instance, the Migdal-Kadanoff
approach is unable to predict the Mean Field exponents of the four
dimensional (ordered) Ising model.

The second method is based on the Mean Field approximation.  This
approach is the classical one that has worked fine in the ordered
Ising Model. Firstly one solves the model in infinite dimensions, then it
is possible to show that the critical properties of the system remain
unchanged up to the so-called upper critical dimension (where the
critical laws are modified by logarithms). Below the upper critical
dimension the thermal fluctuations change the critical behavior, which
can be analyzed using renormalization group techniques. This
approach predicted that the upper critical dimension for spin glasses
is six.

Our main goal is to calculate the logarithmic corrections that predicts
the last approach (i.e. continuous formulation of the problem plus
renormalization group) in six dimensions for the spin glass (its upper
critical dimension) and to compare them with the logarithms found by
Wang and Young~\cite{WAYO93} simulating the six dimensional spin
glass.

Wang and Young performed extensive numerical simulations~\cite{WAYO93}
in order to check whether six was really the upper critical dimension
of the theory~\cite{FISHSOM}. They found the Mean Field critical exponents
($\gamma=1$ and $\nu=1/2$) but they also found logarithmic
corrections, for example, looking at the finite size effects on the
non linear susceptibility. Obviously to close this still open problem
(i.e. whether six is really the upper critical dimension) it is mandatory
to known whether the logarithmic corrections found by Wang and Young
are those predicted by the theory.

Another point of interest is to check that at least when $\epsilon=0$
the approach has a predictive power. The convergence of the
$\epsilon$-expansion is really poor for the $\phi^3$ theory (see
\cite{PERCO4} for an example of this poor convergence in
percolation). In particular for a one component spin glass it is
impossible to re-sum (in the Borel sense) the series for the critical
exponents because all the known terms of the series has the same
sign. This is why the Field Theory approach has not had a great
success. But in this note we show that the underlying approach is
indeed right: it predicts the right logarithmic corrections that has
been found with the computer.

We calculate analytically in the present paper the logarithmic
corrections, and we compare them with those seen by Wang
and Young, finding a very good agreement.

The calculation of the logarithmic corrections (for the correlation
length, the non linear susceptibility and the specific heat) has been
done using two different analytical starting points:
\begin{enumerate}

\item The renormalization group recursion formulas, found by Harris,
Lubensky and Chen~\cite{HALUBCHEN} in the framework of the Wilson
renormalization group~\cite{WILSON}.  In this case we have obtained
the logarithmic corrections for the $m$-component spin glass.

\item The results of De Alcantara et al.~\cite{ALKITMC80,ALKITMC81},
obtained using a Field Theory approach~\cite{AMIT}. In this case the
coefficients, tensor, of the $\phi^3$ term in the actions are
completely general and so we can distinguish our final formula to the
following cases: percolation, Lee-Yang singularities and $m$-component
spin glass (in this case using the results of \cite{GREEN})~\footnote{
Obviously in these works~\cite{ALKITMC80,ALKITMC81,AMIT_P} there is
information about the Potts model, but in six dimensions the Potts
Model (with more than two states) shows a first order phase
transition.}.

\end{enumerate}
Of course, at the end we will get two predictions (but not fully
independent) for the logarithmic corrections for the spin glass case,
that agree between them: thus we have checked that the final formulas
are right (for the spin glass at least). Moreover we have extended the
computation to two other systems namely the percolation and the
Lee-Yang singularities.

Another check of our calculation was done by comparing the logarithmic
correction of the mean cluster size (a susceptibility) and the
logarithm of the correlation length (in $p-p_c$) that we have found in
percolation with the results of the reference ~\cite{PERC_FT}, where
they were computed using a Field Theory approach, the agreement being
perfect. Moreover we have compared the logarithmic corrections to the
critical behavior of the non linear susceptibility for the six
dimensional spin glass with the results of reference ~\cite{FISCH}
obtaining, again, the same formula for the non linear susceptibility.

We remark that we have extended the computation of the logarithmic
corrections for percolation and spin glasses to another set of
observables and, what is the main issue of the paper, we have computed
the cited correction as a function of lattice size. The study of the
six dimensional percolation and the six dimensional Ising spin glass
using series expansions can be seen in references ~\cite{PERC6} and
~\cite{KLEIN} respectively.

The plan of the paper is the following. In the next section we write
down the analytical set-up that we need in the rest of the paper: the
renormalization group recursions of Harris et al., the Field Theory
results of De Alcantara et al. and some useful Mean Field results. In
sections 3 and 4 we deduce using the Wilson Renormalization Group
(WRG) the logarithmic scaling correction as a function of temperature
and lattice size respectively for the $m$-component Ising spin glass.
In section 5 we generalize the previous results to percolation and
Lee-Yang singularities (in temperature and lattice size for a $\phi^3$
theory with imaginary coupling) using the mapping between the results
of Field Theory (FT) and WRG checking that for the Ising spin glass we
recover the results of sections 3 and 4. In section 6 we have computed
the singular part of the free energy for a $\phi^3$ theory with
imaginary coupling just at criticality and in presence of a magnetic
field (LY singularities \cite{FISHERA}) 
and we have compared this result with that of
lattice animals in eight dimension (in this case as a function of the
fugacity, that plays the role of the magnetic field in LY
singularities) obtaining the same result, a further test that the
mapping (a perturbative mapping) 
suggested by Parisi and Sourlas \cite{PASOUR} works even in presence of
logarithms.  Finally in section 7 we present the conclusions. 

\section{Analytical set-up}

In this section we will write the recursion formulas for the
$m$-component spin glass found by Harris et al. \cite{HALUBCHEN}
using renormalization group {\em \`a la} Wilson and the Field
Theoretical renormalization group formulas obtained by De Alcantara et
al. \cite{ALKITMC80,ALKITMC81}.

Moreover we will write down some useful formulas in the Mean Field
framework.

\subsection{Wilson renormalization group (WRG) equations}

One can obtain with  the replica trick and assuming that the replica
symmetry has not been broken, the following starting action for the
$m$-component spin glass
\be
\label{action}
S=\int \d^d x \left[ \frac{1}{2} (\partial_i \phi)^2 + \frac{1}{4} m r \phi^2
-w (n-2) \phi^3 \right] \ .
\ee
where $n$ is the number of replicas.

Harris, Lubensky and Chen~\cite{HALUBCHEN} 
found in a renormalization group calculation {\em \`a la}
Wilson~\cite{WILSON} the following recursion relations ($b$ is
the  scaling factor) for the action (\ref{action})
\begin{eqnarray}\nonumber
r^\prime &=& b^{2-\gamma} \left( r -36(n-2) m w^2 [A(0)-2 K_6 r] 
\log b \right)\ ,\\
w^\prime &=& b^{\epsilon/2-3 \gamma/2} \left(w +36[(n-3)m+1] w^3 K_6
\log b \right)\ , \label{rg_eq1}
\end{eqnarray}
with $\gamma=\gamma(w)=12(n-2) m w^2 K_6$, 
and $\epsilon=6-d$. $A(0)$ and $K_6$ are  constants.

In the spin glass case we assume  we take the replica trick
limit ($n \to 0$) and the number of dimensions to be six
($\epsilon=0$).

We can write Eqs. (\ref{rg_eq1}) in a differential form, by
performing a differential dilatation, obtaining
\begin{eqnarray}\nonumber
\frac{\d r}{\d \log b} &=& (2 -120 K_6 m w^2) r +72 m w^2 A(0)\ ,\\
\frac{\d w}{\d \log b} &=& -36 (2m-1) K_6 w^3 \ .
\label{rg_eq}
\end{eqnarray}
We denote $\beta_{\rm W}(w)\equiv\d w/\d \log b$.

Defining  $t \equiv  r + 36 A(0) m w^2$ we recast the first
equation of (\ref{rg_eq}) in the standard form 
\be
\frac{\d t}{\d \log b} = (2 -120 K_6 m w^2) t \ .
\ee 
The solutions of the WRG equations (Eqs. (\ref{rg_eq}) are
\begin{eqnarray}
\log \frac{t(b)}{t_0} &=& 2 \log b -\frac{5 m}{3 (2m-1)} \log\left(1 +72
w^2(1) K_6 (2m-1) \log b\right )\ , \\ \label{w_sol}
w^2(b)&=& \frac{w^2(1)}{1+72 w^2(1) K_6 (2m-1) \log b} \ , 
\label{sol_rg_eq}
\end{eqnarray}
where $t_0 \equiv t(b=1)$. We are interested
in the asymptotic behavior that reads
\begin{eqnarray}
t(b)& \sim & t_0 b^2 ~(\log b) ^{-5m/(3(2m-1))} \ , \\
w(b)& \sim & \frac{1}{\sqrt{72 (2m -1) K_6}}~ (\log b)^{-1/2} \ . 
\end{eqnarray}

In order to link the previous formulas with the Field Theory approach
we recall
\be
\gamma(w) = -24  m w^2 K_6 \ ,
\label{def_gamma}
\ee
and define $\overline\gamma$ as
\be
\frac{\d t}{\d \log b}=t (2 +\overline \gamma(w)) \ ,
\ee
obtaining
\be
\overline{\gamma}= -120 K_6 m w^2 \ .
\ee

Finally we can write, in this approach, 
the expression of the critical exponents as a
function of $\eta(w)$ and $\overline\gamma(w)$ at the fixed point $w^*$
(where $\beta_{\rm W}(w^*)=0$)
\begin{eqnarray} \nonumber
\nu &=& \frac{1}{2 +\overline{\gamma}(w^*)} \ , \\
\eta &=& \gamma(w^*) \ .
\end{eqnarray}

\subsection{Field Theory Formulas}
\label{ft}
Taking the limit, $\epsilon \to 0$ in the formulas of 
references~\cite{ALKITMC80,ALKITMC81} it is possible to write
 in the notation of Amit's
book~\cite{AMIT} \footnote{ 
        We have recast all the 
        formulas of these references~\cite{ALKITMC80,ALKITMC81}
        to the notation of the Amit's 
        book~\cite{AMIT}. For $\beta(w)$ and $\gamma_\phi(w)$ there are 
        no changes. The only difference is on ${\overline \gamma}_{\phi^2}$. 
        The rule is: ${\overline \gamma}_{\phi^2}=-\gamma_{\phi^2}({\rm
        REFERENCES})$. Here for $\gamma_{\phi^2}({\rm
        REFERENCES})$ we mean the value of $\gamma_{\phi^2}$ found 
        in~\cite{ALKITMC80,ALKITMC81}.}   

\begin{eqnarray} \nonumber
\beta(w)&=&\left(\frac{1}{4} \alpha -\beta \right) w^3 \ , \\ \nonumber
\gamma_\phi(w)&=&\frac{1}{6} \alpha w^2 \ , \\ \nonumber
 {\overline \gamma}_{\phi^2}(w)&=&-\alpha w^2 \ , \\
\gamma_{\phi^2} &\equiv& {\overline \gamma}_{\phi^2} +\gamma_{\phi}
=-\frac{5}{6} \alpha w^2 \ .
\label{RG_FORMULAS}
\end{eqnarray}
The values for $\alpha$ and $\beta$ for different models are written
in Table (\ref{table:param}). We have taken the $\alpha$ and $\beta$  
values from
references \cite{ALKITMC80,ALKITMC81} (percolation and Lee-Yang
singularities) and \cite{GREEN} (spin glasses).

\begin{table}
\centering
\begin{tabular}{|c|c|c|c|} \hline
         &PERC   &  $m$-SG  & LY-S \\ \hline \hline
$\alpha$ &$-1$   & $-4m$     & $-1$\\ \hline 
$\beta$  &$-2$   & $1-3m$    & $-1$\\ \hline 
\end{tabular}
\caption{Values of $\alpha$ and $\beta$ for percolation (PERC),
$m$-component vector spin glass ($m$-SG) and Lee-Yang singularities
(LY-S).}
\protect\label{table:param}
\end{table}

Using the spin glass values for $\alpha$ and $\beta$ it is possible to
link the WRG formulas and the FT ones. Taking $K_6=1/36$  we
found that $\beta_{\rm W} \to -\beta$, $\gamma_\phi \to \gamma$ and
$\overline{\gamma} \to -\gamma_{\phi^2}$. This mapping can be checked 
with the formulas for the critical exponents~\footnote{ The critical
exponents in  FT are~\cite{AMIT}: $\eta=\gamma_{\phi}(w^*)$ and
$\nu=1/(2 -\gamma_{\phi^2}(w^*))$. }.

Finally, it is possible to write a Callan-Symanzik like formula for
the inverse of the susceptibility (see, for instance, reference~\cite{AMIT})
\be
\left[\kappa \frac{\partial}{\partial \kappa} +\beta(u) 
\frac{\partial}{\partial u} -\eta(u)
-\theta(u)\frac{\partial} {\partial t} \right]\chi_R^{-1}(t,u,\kappa)=0 \ .
\label{CS_SUS}
\ee
where $\theta(u)\equiv -(\overline{\gamma}_{\phi^2}(u) +\gamma_{\phi}(u))$;
$\kappa$ is the momentum scale, $u$ is the dimensionless renormalized 
coupling and $t$ is the
renormalized reduced temperature. 

\subsection{Mean Field}

One subtle point in the present calculation is the presence of
irrelevant dangerous variables . 
To understand when and how they
appear we need to analyze the theory in the Mean Field framework.

The free energy is
$$
F(r_0,w_0)=\frac{r_0}{2} M^2 +\frac{w_0}{3!} M^3 \ .
$$
If $r_0>0$ the only solution that minimizes the free energy is $M=0$. But
if $r_0<0$ the solution is $M=2|r_0|/w_0$, and the free
energy at this minimum is
$$ 
F_{\rm min}=-\frac{2}{3} \frac{|r_0|^3}{w_0^2} \ .
\nonumber
$$
The specific heat is the second derivative of $F_{\rm min}$ with
respect $r_0$:
\be     
C \propto \frac{|r_0|}{w_0^2} \ ,
\label{mf_cs}
\ee
i.e. $\alpha=-1$, and $w_0$ is a
dangerous variable~\footnote{
        Mean Field also predicts that the specific heat vanishes
        in the whole paramagnetic phase.}. $w_0$ is dangerous because
the RG prediction is  that $w_0 \to 0$ for larger blocking and it appears
in the denominator in the free energy expression.

It is easy to obtain that
\be
\chi^{-1} \propto r_0 \ ,
\label{mf_sus}
\ee
i.e. $\gamma=1$ and thereby for this observable 
we find that it does not depend  on $w_0$.

Finally we remark that in the Mean Field approximation we have  
$\chi(r_0)\propto \xi(r_0)^2$: i.e. $\nu=1/2$.

\section{Logarithmic corrections in temperature (WRG) for spin glasses}

The starting point is the usual formula for the propagator 
at momentum $\bf k$:
$$
G({\bf k},r_0,w_0,\Lambda)=b^2 \zeta(b) G(b ~{\bf k}, r(b), w(b),
\Lambda) \ ,
$$
where $\zeta(b)$ is defined as
\be
\frac{\d \log \zeta(b)}{\d \log b} \equiv -\gamma(w(b)) \ ,
\label{zeta}
\ee
$\gamma(w)$ was defined in Eq. (\ref{def_gamma}) and $\Lambda$ is
the cut-off (see reference \cite{LE_BELLAC} for more details). Solving
Eq. (\ref{zeta}) we obtain
$$
\zeta(b)\simeq (\log b)^{m/(3(2m-1))} \ .
$$
The susceptibility is nothing but the propagator at zero momentum, and
so
\be 
\chi(r_0,w_0,\Lambda)=b^2 \zeta(b) \chi(r(b), w(b), \Lambda) \ .
\label{susceptibility}
\ee

An equivalent approach  is to start from the following formula for the
singular part of the free energy (\cite{SALAS_SOKAL})
$$
f_{\rm sing}(r_0,w_0,h_0)=b^{-6} f_{\rm sing}(r(b),w(b),h(b)) \ ,
\label{f_sing}
$$
where $h(b)$ is the re-scaled magnetic field that satisfies the
following recursion formula
$$
h(b)=b^{d_h} h_0 \ ,
$$
with $d_h=(d-\gamma(w))/2+1$.

The susceptibility reads
$$
\chi \propto \left. \frac{\partial^2 f_{\rm sing}}{\partial h_0^2} 
\right|_{h_0=0} \ ,
$$
obtaining again Eq. (\ref{susceptibility}). 

Taking, as usual, a $b^*$-value such that $t(b^*)=1$, i.e.,
$$
b^* \propto t_0^{-1/2} ~(\log t_0) ^{5m/(6(2m-1))} \ ,
$$
we obtain
\be
\chi \propto t_0^{-1}~ (\log t_0)^{2m/(2m-1)}.
\ee

The correlation length verifies
\be
\xi(r_0,w_0,\Lambda)=b~ \xi(r(b), w(b), \Lambda)
\ee
and therefore
\be
\xi \propto b^* = t_0^{-1/2}~ (\log t_0) ^{5m/(6(2m-1))}
\ee

We remark that the remaining factors in the above deduction (i.e. $
\xi(1, w(b), \Lambda)$ and $\chi(1,w(b), \Lambda)$) do not diverge
(see Eq. (\ref{mf_sus})): they are the correlation and the
susceptibility far away of the critical point; and so computed in the
Mean Field approximation. We have also found above that in the Mean
Field Calculation $\xi$ and $\chi$ do not depend on $w$.

We can finally write down the formulas for the Ising spin glass ($m=1$):
\begin{eqnarray} \nonumber
\chi &\propto & t_0^{-1}~ (\log t_0)^2 \ , \\
\xi &\propto & t_0^{-1/2}~ (\log t_0) ^{5/6} \ .
\end{eqnarray}
At this point, we can compare with the analytical result of Fisch and
Harris ~\cite{FISCH} where it was found $t_0^{-1}~ (\log t_0)^2$, and so
we use their result for the susceptibility as check of our calculation.

In general the susceptibility verifies 
 $\chi=\xi^{2-\eta}$. In the $\phi^4$ theory,
in four dimensions,$\phi^4_4$, it is clear that $\eta=0$ and 
$\chi=\xi^2$, while
in $\phi^3$ in six dimension (which we denote as $\phi^3_6$),
 we have again that $\eta=0$ but with 
induced logarithmic corrections and so $\chi \neq \xi^2$. This fact is
related with the fact that in $\phi^4_4$, $\zeta={\rm Constant}$, while in the
$\phi^3_6$ this does not hold.

For the specific heat we use again the expression  the singular part
of the free energy (Eq. (\ref{f_sing})).
The singular part of the specific heat is
$$
C \propto \left. \frac{\d^2 f_{\rm sing}}{\d t_0^2} \right|_{h_0=0} \ .
$$

We choose again the same $b^*$ as above and we can finally write
\be
C \propto t_0 (\log t_0)^{-\frac{3m+1}{2m -1}} \ ,
\ee
where we have used that in the Mean Field
approximation the specific heat behaves like $C \propto 1/w(b)^2$ 
(see Eq. (\ref{mf_cs})). 
In particular, fixing $m=1$, one gets
\be
C \propto t_0 (\log t_0)^{-4} \ .
\ee

\section{Finite Size Scaling formulas with logarithmic corrections (WRG) for 
spin glasses
}

The scaling of the singular part of the free
energy in the presence of a magnetic field, $h_0$, is (in six dimensions)
\cite{SALAS_SOKAL}
\begin{equation}
\label{main_rg}
f_{\rm{sing}}\left(r_0,w_0,h_0,\frac{1}{L}\right) 
=b^{-6} f_{\rm{sing}}\left(r(b),w(b),h(b),\frac{b}{L}\right) \ ,
\end{equation}
where we have introduced a new coupling, the system size $L$, which
scales trivially with a RG transformation ($1/L \to b/L$). 
As usually the magnetic field verifies~\cite{LE_BELLAC}:
\begin{equation}
\frac{\d \log h(b)}{\d \log b} =\frac{d}{2}+1
-\frac{\gamma(w)}{2}\ .
\label{RGH}
\end{equation}

In the asymptotic regime, the solution of (\ref{RGH}) is
$$
h(b)= h_0 b^4 (\log b)^{\frac{m}{6(2m-1)}}\ .
$$

Performing a RG transformation with $b=L$,  we keep just one degree of
freedom (see ref.~\cite{BLOTE} for more details). The free
energy of this system reads
$$
f(r',w',h',L=1) \equiv
\displaystyle\log \int \d \phi ~
\displaystyle\exp\left\{-\left[\frac{r'}{4}  \phi^2 
- h'
\phi -(n- 2) w^\prime 
 \phi^3  
  \right]\right\}\ .
$$
Using the standard approach ~\cite{BLOTE} 
we re-scale the $\phi$ variable by means of
$\phi^\prime=w^{\prime 1/3} \phi$. The free energy can be written as
\begin{equation}
\label{second_rg}
f(r',w',h',L=1)= 
{\hat f}\left(\frac{r'}{w'^{2/3}},\frac{h'}{w'^{1/3}}\right)\ ,
\end{equation}
obtaining finally
\begin{equation}
f_{\rm{sing}}\left(r_0,w_0,h_0,\frac{1}{L}\right)=L^{-6} 
{\hat
f}\left(\frac{r(L)}{w(L)^{2/3}},\frac{h(L)}{w(L)^{1/3}}\right)\ .
\label{fun_fin}
\end{equation}
This formula also holds for a generic $\phi^3$.

We remark that $w$  is a {\em dangerous} (marginally)
irrelevant variable ~\cite{SOKAL2,FISHER} and we need to do with care
all the analytical steps (it is not correct to substitute $w$ for its
asymptotically value, $w=0$, because the free energy depends on
inverse powers of $w$).

To compute the thermodynamical quantities in the critical region one
just need to take the appropriate derivatives of $f_{\rm sing}$.
In order to compute the logarithmic corrections it will prove useful
take into account 
Eqs. (\ref{w_sol}) and (\ref{main_rg}), and the
following Taylor expansion 
(which depends on $r(L)=t(L)-36 A(0) m w(L)^2$
and that  $t_0=0$ implies  $t(L)=0$)
\begin{equation}
\begin{array}{rcl}
\left. \partial^2_i{\hat f}(r(L)/w(L)^{2/3},0)\right|_{t_0=0}&=&
\partial^2_i{\hat f}(-36 A(0) m w(L)^{4/3},0) \\
\\
&= & \displaystyle\partial^2_i{\hat f}(0,0) +O(w(L)^{4/3}) \ ,
\end{array}
\label{taylor}
\end{equation}
where $\partial_i$ is the partial derivative with respect to the
$i$-th argument.

The equation (\ref{taylor}) gives us the leading correction to the scaling: it
is just the term $O(w(L)^{4/3})$ that modifies the scaling behavior given
by $\partial^2_i{\hat f}(0,0)$. In a $m$-component spin glass we
should expect correction to scaling proportional to $1/(\log L)^{2/3}$.

As we are interested in the behavior with the
lattice size just at the infinite volume critical temperature
($t_0=0$) , the
susceptibility can be written as
\begin{eqnarray}\nonumber
\chi &\propto& \left. \frac{\partial^2 f_{\rm sing}}
{\partial h_0^2}\right|_{h_0=t_0=0}
=L^{-6} 
\left(\frac{\partial h(L)}{\partial h_0}\right)^2 \left. 
\frac{\partial^2}{\partial h(L)^2}{\hat
f}\left(r(L)/w(L)^{2/3},h(L)/w(L)^{1/3} \right)
\right|_{h_0=t_0=0} \\
&\propto& L^2 (\log L)^{\frac{3m-1}{3 (2m-1)}} \left[1 +
\frac{A}{(\log L)^{2/3} } \right]  \ ,
\label{sus}
\end{eqnarray}
where $A$ is a  constant.

At this point we can compare our prediction for the logarithmic
correction for one component spin glass 
\be
\chi \propto L^2 (\log L)^{\frac{2}{3}} \ ,
\ee
with which was found in numerical simulations by Wang and Young~\cite{WAYO93}:
$$
\chi \propto L^2 (\log L)^{0.64} \ .
$$
The agreement is very good.

The specific heat can be computed analogously
\begin{eqnarray}\nonumber
C &\propto& \left. \frac{\partial^2 f_{\rm sing}}
{\partial t_0^2}\right|_{h_0=t_0=0} 
=L^{-6} 
\left(\frac{\partial r(L)}{\partial t_0}\right)^2
\left.\frac{\partial^2}{\partial r^2} {\hat
f}(r(L)/w(L)^{2/3},h(L)/w(L)^{1/3})\right|_{h_0=t_0=0} \\
&\propto&  L^{-2} (\log L)^{-\frac{2(3m+1)}{3(2m-1)}}\ .
\end{eqnarray}

At zero magnetic field, the correlation length scales as
\begin{equation}
\left. \xi(r_0,w_0,1/L) \right|_{t_0=0}=L \left. \xi(r(L),w(L),1) 
\right|_{t_0=0} \ ,
\label{XI1}
\end{equation}
where $\xi(r(L),w(L),1)$ must be evaluated with the free energy
(\ref{second_rg}). Consequently, the mass squared term is
$$
\left( \left.\xi(r,w,1)\right|_{t_0=0}\right)^{-2}
= \left. \frac{r(L)}{w(L)^{2/3}}
\right|_{t_0=0} \propto w(L)^{4/3} \ ,
$$
and so
\begin{equation}
\xi(r_0,u_0,1/L) \propto L w(L)^{-2/3} \propto L (\log L)^{\frac{1}{3}} \ .
\end{equation}
The independence of the logarithmic corrections of the correlation
length on the number of components for spin glasses is similar to the
$\phi^4_4$ case, where no dependences on the number of components was
found in the exponent of the logarithmic corrections \cite{BREZIN}.

Finally we can also compute the shift of the apparent critical
temperature. It can be defined  as the temperature where the
susceptibility (or specific heat)  measured in a finite volume 
shows a maximum. Using the
formula (\ref{sus}) for the susceptibility
without imposing 
$t_0=0$  we obtain
$$
\chi \propto L^2 (\log L)^{\frac{2m}{3 (2m-1)}} 
\partial^2_2{\hat f}(r(L)/w(L)^{2/3},0)\ .
$$
The maximum of $\chi$ as a function of $L$ and $t$ is not just at
$t_0=0$, but it is fixed by the condition  
$$
r(L)/w(L)^{2/3}=\frac{t(L) - 36 A(0) m  w(L)^2}{w(L)^{2/3}}=x_{\rm max} \ ,
$$
i.e. the function
$\partial^2_2{\hat f}(x,0)$ has a maximum at $x=x_{\rm max}$ .

As  \hbox{$t \propto T_{\rm c}(\infty)-T_{\rm c}(L)$}, it follows that
\begin{equation}
T_{\rm c}(\infty)-T_{\rm c}(L) \propto L^{-2} 
\left(\log L \right)^{\frac{3m +1 }{3 (2m -1)}} \ .
\label{shift}
\end{equation} 

\section{Percolation and Lee-Yang Singularities Formulas}

Following the procedure used in the two previous sections we can write
general formulas, not just for the spin glass case as we have done in
the previous part of the paper. Thus, by using the mapping between the
WRG formulas ($\beta_{\rm W}(w), \gamma(w)$ and ${\overline
\gamma}(w)$) and the FT ones ($\beta(w), \gamma_{\phi}(w)$ and
$\gamma_{\phi^2}(w)$) we can obtain the formulas for percolation and
Lee-Yang singularities.

The starting point are  the following general formulas
\begin{eqnarray}\nonumber
\frac{\d \log h}{\d \log b} &=& \frac{d+2}{2} -\frac{\gamma(\omega)}{2}\ ,\\
\nonumber
\frac{\d w}{\d \log b} &=& \beta_{\rm W}(w) \ ,\\
\frac{\d t}{\d \log b} &=& (2 +{\overline \gamma}(\omega)) t \  ,
\label{GF}
\end{eqnarray}
and so
\begin{eqnarray}\nonumber
\int_{w_0}^{w(b)} \frac{\d w}{\beta_{\rm W}(w)} &=& \log b \ ,\\ \nonumber
h(b) &=& h_0 b^{(d+2)/2} 
\exp\left[-\frac{1}{2} \int_{w_0}^{w(b)} \d w~ 
\frac{\gamma(w)}{\beta_{\rm W}(w)} \right] \ , \\
t(b) &=& t_0 b^2 \exp\left[ \int_{w_0}^{w(b)} 
\d w~\frac{ {\overline{\gamma}(w)}}{\beta_{\rm W}(w)} \right] \ .
\end{eqnarray}
where $t_0 \equiv t(b=1)$, $h_0 \equiv h(b=1)$ and $w_0\equiv w(b=1)$.

We rewrite the general formula for the singular part of the free energy
$$
f_{\rm sing}(r_0,w_0,h_0)=b^{-d} f_{\rm sing}(r(b),w(b),h(b)) \ .
$$
We define $b^*$ such that $t(b^*)=1$, and so $b^*$ is a function of $t_0$.

Taking two derivatives with respect to the magnetic field on the
singular part of the free energy we obtain the formula for the
susceptibility~\footnote{ In the case of  percolation we identify
the susceptibility with the mean cluster size~\cite{STAUFFER}.}
\be
\chi\propto (b^*)^2 \exp\left[-\int_{w_0}^{w(b^*)} 
\d w ~\frac{ \gamma(w)}{\beta_{\rm W}(w)} \right] \ .
\ee
This formula is valid in any dimension and theory.

The specific heat is~\footnote{ We identify, for  percolation, the
specific heat with the second derivative of the singular part of the
total number of clusters, $M_0$, with respect to the dilution
~\cite{STAUFFER}. In the percolation
case the temperature is identified with the probability $p$, and so the
reduced temperature means $p-p_c$.}
\be
C \propto (b^*)^{4-d} \frac{1}{w^2(b^*)} \exp\left[2 \int_{w_0}^{w(b^*)} 
 \d w~\frac{ {\overline{\gamma}(w)}}{\beta_{\rm W}(w)} \right]   \ . 
\ee
This formula is valid in any dimension and only for a $\phi^3$ theory
(we have used that in a $\phi^3$ theory the specific heat far from the
critical point behaves as $1/w^2$; in a $\phi^4$ theory we should change
the factor $1/w^2$ to $1/w$).

Finally the correlation length is 
\be
\xi \propto b^* \ .
\ee

The Finite Size Scaling Formulas are
\be
\chi\propto L^2 \frac{1}{w(L)^{2/3}} \exp\left[-\int_{w_0}^{w(L)}  
\d w ~\frac{ \gamma(w)}{\beta_{\rm W}(w)} \right]  \ ,
\ee
for the susceptibility. The specific heat is
\be
C \propto L^{4-d} \frac{1}{w(L)^{4/3}}  \exp\left[2 \int_{w_0}^{w(L)} 
 \d w~\frac{ {\overline{\gamma}(w)}}{\beta_{\rm W}(w)} \right]
 \ , 
\ee
the correlation length is given by
\be
\xi \propto L w(L)^{-2/3} \ , 
\ee
and the shift of the critical temperature is
\be
T_c(L)-T_c(\infty) \propto L^{-2} w(L)^{2/3}  \exp\left[- \int_{w_0}^{w(L)} 
 \d w~\frac{ {\overline{\gamma}(w)}}{\beta_{\rm W}(w)} \right] 
 \ .
\ee
The FSS formulas for the specific heat, susceptibility, correlation
length and shift are valid in any dimension but only for a  $\phi^3$
theory. For $d<6$ we have $\lim_{L\to \infty} w(L)=w^* \neq 0$ and 
we get the standard (i.e. without logs corrections) Finite Size scaling 
formulas.

Using the mapping between $\beta_{\rm W}(w), \gamma(w)$ and
$\overline{\gamma}(w)$ and $\beta(w), \gamma_\phi(w)$ and
$\gamma_{\phi^2}(w)$ of section \ref{ft} we can write:
\begin{eqnarray}
\chi &\propto & t_0^{-1} \label{chi_fin}
\left[ \log t_0 \right]^{\frac{2 \alpha}{4 \beta- \alpha}} \ , \\
 C   &\propto & t_0 
\left[ \log t_0 \right]^{-\frac{6 \alpha-4 \beta}{4 \beta- \alpha}} \
, \\ \label{xi_fin}
\xi  &\propto & t_0^{-1/2}
\left[ \log t_0 \right]^{\frac{5 \alpha}{6(4 \beta- \alpha)} }\ . 
\end{eqnarray}

In Table \ref{table:beta} we report the exponent of the logarithm 
observables for percolation, spin glasses and Lee-Yang singularities.

The results $\chi \propto t_0^{-1} [\log t_0]^{2/7}$ and
$\xi \propto t_0^{-1/2} [\log t_0]^{5/42}$ for percolation
were found solving a Callan-Symanzik equation in reference
\cite{PERC_FT}, and therefore we use these results as check of the
above calculation.

\begin{table}
\centering
\begin{tabular}{|c|c|c|c|} \hline
 &PERC   &  $m$-SG  &  LY-S \\ \hline \hline
$\chi$ &$2/7$    & $2m/(2m-1)$        & $2/3$\\ \hline 
$C$    &$2/7$    & $-(1+3m)/(2m-1)$  & $-2/3$\\ \hline 
$\xi$  &$5/42$   & $5 m/(6(2m-1))$    & $5/18$\\ \hline 
\end{tabular}
\caption{Values of the exponent of the logarithmic correction, in
reduced temperature, for
the susceptibility, specific heat and correlation length.}
\label{table:beta}
\end{table}

As a function of the lattice size we found the following formulas.
\begin{eqnarray}
\chi(L,t_0=0) &\propto& L^2
\left[ \log L \right]^{\frac{4 \beta}{3(4 \beta- \alpha)}} \ , \\
 C (L,t_0=0)  &\propto& L^{-2}
\left[ \log L \right]^{-\frac{12 \alpha -8 \beta}{3(4 \beta- \alpha)}} \ , \\
\xi(L,t_0=0)  &\propto& L \left[ \log L \right]^{1/3}\ , \\ 
\Delta T_c &\propto& L^{-1/2}
\left[ \log L \right]^{-\frac{4 \beta - 6 \alpha}{3(4 \beta- \alpha)}} \ . 
\end{eqnarray}

In Table \ref{table:lattice} we report the exponent of the logarithm 
(in lattice size) for different  models.

\begin{table}
\centering
\begin{tabular}{|c|c|c|c|} \hline
         &PERC&  $m$-SG     & LY-S \\ \hline \hline
$\chi$ &$8/21$& $(3m-1)/(3(2m-1))$        & $4/9$\\ \hline 
$C$    &$4/21$& $-2(1+3m)/(3(2m-1))$   & $-4/9$\\ \hline 
$\xi$  &$1/3$ & $1/3$   & $1/3$\\\hline 
$\Delta T$&$-2/21$& $(1+3m)/(3(2m-1))$  & $2/9$\\ \hline 
\end{tabular}
\caption{Values of the exponent of the logarithmic correction, in
lattice size, for
the susceptibility, specific heat and correlation length. We also show
the logarithm factor for
 shift of the apparent critical temperature with the lattice size.}
\label{table:lattice}
\end{table}

Moreover the leading correction to the scaling, for all the models
described in this paper (in general for all the models described by a
generalized $\phi^3$), is proportional to $1/(\log L)^{2/3}$.

Finally, we can re-obtain part of the previous results (in temperature) using
standard Field Theoretical techniques.

The starting point is the solution of the Callan-Symanzik
like equation for the susceptibility (see Eq. (\ref{CS_SUS})). 
The solution of this equation is~\cite{AMIT}
$$
\chi_R^{-1}(r_0, w) \propto r_0 \exp\left[- \int_1^{r_0}
\left(\overline{\eta}(w(x))+ \overline{\theta}(w(x))\right) 
\frac{\d x}{x} \right] \ ,
$$
where 
$$
\overline{\eta}(w)\equiv \frac{\gamma_\phi(w)}{2-\gamma_{\phi^2}(w)}\ ,
$$
$$
\overline{\theta}(w)\equiv
-\frac{\gamma_{\phi^2}(w)}{2-\gamma_{\phi^2}(w)} \ ,
$$
and $w(x)$ verifies
$$
\frac{\d w}{\d \log x}=
\overline{\beta}(w)\equiv \frac{\beta(w)}{2-\gamma_{\phi^2}(w)} \ , 
$$
with the initial condition $w(x=1)=\hat w_0$.

Using the formulas (\ref{RG_FORMULAS}) of the FT approach we find 
$$
\chi_R^{-1}(r_0) \propto r_0 ~(\log r_0)^{2 \alpha/(\alpha-4\beta)} \ .
$$
And we find again the same law (see Eq. (\ref{chi_fin})).

We can repeat the above calculation for $\xi^{-2}$. In this case,
solving the correspondent Callan-Symanzik equation for $\xi^{-2}$, we
arrive to the following formula~\cite{AMIT}
$$
\xi^{-2}(r_0, w) \propto r_0 ~\exp\left[- \int_1^{r_0}
\overline{\theta}(w(x)) 
\frac{\d x}{x} \right] \ .
$$
Being the solution
$$
\xi^{-1} \propto r_0^{1/2}~ (\log r_0 )^{5 \alpha/(6 \alpha-24 \beta)}
\ .
$$
Again we have obtained the same result (see Eq. (\ref{xi_fin})).

\section{Lee-Yang singularities and lattice animals}

In this section we will compute the singular part of the free energy for
a $\phi^3$ theory with imaginary coupling at criticality (which
describes the LY singularities \cite{FISHERA}) as a function of the
magnetic field (the results of the preceding section for this model
were as a function of the reduced temperature or lattice size- in
both cases in absence of magnetic field-).

Once that we have this result we will compare it with the formula
found for lattice animals.

M. Fisher  showed \cite{FISHERA} a mapping between the Ising model with
magnetic field in the paramagnetic phase and a $\phi^3$ theory with an
imaginary
coupling  at its critical point. Following the steps described 
above, the first goal is to compute the Mean Field behavior. The free
energy for a $\phi^3$ at its critical point in presence of a magnetic
field $h_0$ is (see reference \cite{FISHERA} for more details)
\begin{equation}
\nonumber
F(r_0=0,w_0,h_0)=h_0 M + \frac{w_0}{3!} M^3 .
\end{equation}
By computing the minimum of the free energy, 
we can write the free energy at this minimum
\begin{equation}
F_{\rm min}\propto \frac{h_0^{3/2}}{w_0^{1/2}} .
\label{free}
\end{equation} 

The next step is to write the renormalization group equation for the
singular part of the free energy at the critical point
\begin{equation}
f_{\rm sing}(r_0=0,w_0,h_0)=b^{-6} f_{\rm sing}(0,w(b),h(b)) . 
\end{equation}
Now, $h_0$ is the relevant parameter and so we choose $b^*$ by means the
relation $h(b^*)=1$, obtaining (by solving the first and second
equations of Eqs. (\ref{GF}))
\begin{equation}
b^*(h_0) \simeq h_0^{-1/4} (\log h_0)^{-1/72} .
\end{equation}
Finally we can write the behavior of the singular part of the free
energy
\begin{equation}
f_{\rm sing} \simeq (b^*)^{-6} \frac{h(b^*)^{3/2}}{w(b^*)^{1/2}} ,
\end{equation}
where we have used Eq. (\ref{free}). Using  $h(b^*)=1$, and the
behavior of $w$ with $b$ we obtain
\begin{equation}
f_{\rm sing} \simeq h_0^{3/2} (\log h_0)^{1/3} .
\label{f_s}
\end{equation}
This is the behavior of the singular part of the free energy of the LY
singularities in six dimensions.  This formula defines the $\sigma$
exponent for the LY singularities by means: 
$f_{\rm sing}= h_0^{\sigma+1}$.
Obviously we have recovered the MF result:  $\sigma=1/2$ but modified by a
logarithmic factor.

If the mapping proposed by Parisi and Sourlas holds then the behavior
of the singular part of the free energy for lattice animals in eight
dimension should be given by Eq.  (\ref{f_s}) (changing the magnetic
field by the fugacity), but this formula is just the formula computed
by Lubensky and Isaacson for the singular part of the free energy for
lattice animals in eight dimensions \cite{LUBENSKY}. Another test of
this formula was done in reference \cite{L_ANIMALS} using series
expansions directly on lattice animals.  

The above calculation (\ref{f_s}) supports again the correctness of
the mapping between lattice animals in $d+2$ dimensions and the LY
singularities in $d$ dimensions. 

\section{Conclusions}

In this paper we have computed the logarithmic corrections for a
generic $\phi^3$ theory at its upper critical dimension both in the
reduced temperature as well as in the size of the system at
criticality. Moreover we have computed the leading corrections to
the scaling and the shift of the apparent critical temperature.

We have distinguish the formulas to the following cases: percolation,
$m$-component spin glasses and Lee-Yang
singularities.

We have compared the results for the one-component spin glass with
the corrections found numerically and the agreement between the theory
and the simulations is very good.

Therefore we believe that the present computation of the logarithmic
corrections for the one component spin glass and the agreement of
these with the numerical simulations strongly support that six is the
upper critical dimension for $m=1$ spin glasses. 

Finally  we have tested the (perturbative) mapping between lattice
animals in eight dimensions and LY singularities in 6 dimensions by
computing the free energy in the LY model.

\section{\protect\label{S_ACKNOWLEDGES}Acknowledgments}

J. J. R.-L. is supported by an EC HMC (ERBFMBICT950429) grant.  We
wish to thank H. G. Ballesteros, L. A. Fern\'andez, V. Mart\'{\i}n
Mayor, A. Mu\~noz Sudupe, G. Parisi and D. Stauffer for interesting
discussions. We also wish to thank the referees for pointing us useful
comments and the link between LY singularities and lattice animals.


\newpage

\begin{thebibliography}{100}

\bibitem{CARDY} J. L. Cardy (ed), {\em Finite-Size Scaling}
(North-Holland, Amsterdam 1988).

\bibitem{BOOK} P. Young (ed.), {\em Spin Glasses and Random Fields}
(World Scientific, Singapore 1997). 

\bibitem{STAUFFER} D. Stauffer and A. Aharony, {\em Introduction to
the Percolation Theory} (Taylor and Francis,London 1994) (Revised
second edition). 

\bibitem{YL} T. D. Lee and C. N. Yang, Phys. Rev. {\bf 87}, 404 (1952).

\bibitem{PARISI} G. Parisi, {\em Field Theory, Disorder and
Simulations} (World Scientific, Singapore 1994). 

\bibitem{WILSON} K. G. Wilson, Rev. Mod. Phys. {\bf 47}, 773
(1975).

\bibitem{AMIT} D. J. Amit, {\em Field Theory, the Renormalization Group,
zinand Critical Phenomena} (World Scientific, Singapore 1984) (Revised
second edition). 
Gen
\bibitem{LE_BELLAC} M. Le Bellac, {\em Quantum and Statistical Field
Theory} (Oxford Science Publications 1991).

\bibitem{ZINN} J. Zinn-Justin, {\em Quantum Field Theory and Critical
Phenomena} (Oxford Science Publications 1990).

\bibitem{BREZIN} E. Br\'e, J. Physique {\bf 43}, 15 (1982).

\bibitem{SALAS_SOKAL} J. Salas and A. Sokal, J. Stat. Phys. 88 (1997)
567

\bibitem{AMIT_P} D. J. Amit, J. Phys. A: Math. . {\bf 9}, 1441 (1976).

\bibitem{ALKITMC80} O. F. de Alcantara, J. E. Kirkham and A. J. McKane, 
J. Phys. A: Math. Gen. {\bf 13}, L247 (1980).

\bibitem{DROPLET} W. L. McMillan, J. Phys. C 17, 3179 (1984);
A. J. Bray and M. A. Moore, in {\em Heidelberg Colloquium on Glassy
Dynamics}, edited by J. L. Hemmen and I. Morgenstern (Springer Verlag,
Heidelberg, 1986), 121; D. S. Fisher and D. A. Huse,
Phys. Rev. Lett. 56, 1601 (1986); Phys. Rev. B 38, 386 (1988). 

\bibitem{NS}
  C. M. Newman and D. L. Stein, Phys. Rev. E. 57, 1356 (1998)
  and references therein.

\bibitem{4DDIS} 
H. G. Ballesteros, L.A. Fern\'andez, V. Martin-Mayor, A. Mu\~noz Sudupe
and G. Parisi, Nucl. Phys. B  512, 681 (1998).

\bibitem{FISHSOM} D. Fisher and H. Sompolinsky, Phys. Rev. Lett., 54,
1063 (1985).   

\bibitem{ALKITMC81} O. F. de Alcantara, J. E. Kirkham and A. J. McKane, 
J. Phys. A: Math. Gen. {\bf 14}, 2391 (1981).

\bibitem{GREEN} J. E. Green, J. Phys. A: Math. Gen. {\bf 17}, L43
(1985).

\bibitem{HALUBCHEN} A. B. Harris, T. C. Lubensky and J-H. Chen,
Phys. Rev. Lett. {\bf 36}, 415 (1976).

\bibitem{PERC_FT} J. W. Essam, D. S. Gaunt and A. J. Guttmann,
J. Phys. A:Math and Gen  11, 1983 (1978).

\bibitem{PERC6} J. Adler, Y. Meir, A. Aharony and A. B. Harris,
Phys. Rev. B, 41, 9183 (1990).

\bibitem{FISCH} R. Fisch and A. B. Harris. Phys. Rev. Lett. 38, 785 (1977). 

\bibitem{KLEIN} L. Klein et al. Phys. Rev. B 43 11249 (1991).

\bibitem{LUITBLO} E. Luijten and W. J. Bl\"ote, Phys. Rev. Lett. {\bf
76}, 1557 (1996); Erratum  {\bf 76}, 3662 (1996).

\bibitem{WAYO93} J. Wang and A. P. Young, J. Phys. A: Math. Gen. {\bf
26}, 1063 (1993).

\bibitem{PERCO4} H. G. Ballesteros, L.A. Fern\'andez, V. Martin-Mayor,
A. Mu\~noz Sudupe, G. Parisi and J. J. Ruiz-Lorenzo, 
Phys. Lett. B {\bf 400}, 346 (1997).

\bibitem{BLOTE} E. Luijten and W. J. Bl\"ote,  Phys. Rev. Lett. {\bf
76}, 1557 (1996); Erratum  {\bf 76} (1996) 3662.

\bibitem{SOKAL2} A. C. D. van Enter, R. Fern\'andez and A. D. Sokal,
J. Stat. Phys. 72 (1994) 879.

\bibitem{FISHER} M. E. Fisher, in {\em Renormalization Group in
Critical Phenomena and Quantum Field Theory: Proceedings of a
conference}. Temple University ( Philadelphia,
1974).

\bibitem{LUBENSKY} T. C. Lubensky and J. Isaacson,
Phys. Rev. Lett. {\bf 41}, 829 (1978); {\bf 42}, 410(E) (1979);
Phys. Rev. A {\bf 20}, 2130 (1979).

\bibitem{L_ANIMALS} J. Adler et al., Phys. Rev. B {\bf 38}, 4941 (1988).

\bibitem{PASOUR} G. Parisi and N. Sourlas, 
Phys. Rev. Lett. {\bf 46}, 871 (1981).

\bibitem{FISHERA} M. E. Fisher, Phys. Rev. Lett. {\bf 40}, 1610 (1978).

\bibitem{HALU} A. B. Harris and T. C. Lubensky, Phys. Rev. B {\bf 24}, 
2656 (1981).

\end{thebibliography}
\end{document}